ACTUATORS, MICROFLUIDICS, NANOBUBBLES

# New type of microengine using internal combustion of hydrogen and oxygen

Vitaly B. Svetovoy[*,1,2], Remco G. P. Sanders[1], Kechun Ma[1] & Miko C. Elwenspoek[1,3]

[1] MESA+ Institute for Nanotechnology, University of Twente, PO 217, 7500 AE Enschede, The Netherlands (v.svetovoy@utwente.nl)

[2] Institute of Physics and Technology, Yaroslavl Branch, Russian Academy of Sciences, 150007, Yaroslavl, Russia

[3] FRIAS, University of Freiburg, 79104 Freiburg, Germany

**Microsystems become part of everyday life but their application is restricted by lack of strong and fast motors (actuators) converting energy into motion. For example, widespread internal combustion engines cannot be scaled down because combustion reactions are quenched in a small space. Here we present an actuator with the dimensions $100\times100\times5$ μm$^3$ that is using internal combustion of hydrogen and oxygen as part of its working cycle. Water electrolysis driven by short voltage pulses creates an extra pressure of 0.5 – 4 bar for a time of 100 – 400 μs in a chamber closed by a flexible membrane. When the pulses are switched off this pressure is released even faster allowing production of mechanical work in short cycles. We provide arguments that this unexpectedly fast pressure decrease is due to spontaneous combustion of the gases in the chamber. This actuator is the first step to truly microscopic combustion engines.**

The last decennium has witnessed an impressive trend to miniaturize systems of virtually any kind. This trend has many reasons: small systems are often cheaper to produce, they can have properties large systems have not, and they may facilitate use of large systems (cars, for example). An important and generic component in microsystems is the actuator. It plays role of a motor transforming electricity or other kind of energy into mechanical power. In contrast with large scale systems, where effective engines are available (internal combustion or electromagnetic motors), microsystems suffer from lack of strong and fast actuators[1,2]. Small electromagnetic motors cannot generate forces of useful magnitude but internal combustion engines just do not exist at the microscale.

The existing microcombusters[3] cannot be a part of microsystems since they have only one dimension in the range of submillimeters. Microengines using gas combustion perform poorly due to increased heat losses via the volume boundary[4,5] when the volume decreases. However, recently we observed that the reaction between $H_2$ and $O_2$ gases can be ignited spontaneously in stoichiometric nanobubbles smaller than 200 nm overcoming this restriction in some way[6,7]. The mechanism is still not clear, but it is expected that both high Laplace pressure and fast dynamics are important[6]. It is not obvious that the reaction in nanobubbles and performance of the microscopic actuator are related. Nevertheless, we speculate that the gas combustion in the chamber happens via combustion in transitional nanobubbles. Nanobubbles already demonstrated counterintuitive properties such as unexpectedly long life time of the surface nanobubbles (see[8] for a review). Mechanism of this stability is still debated[9-12] after more than 10 years of active discussion.

Existing microactuators are using mostly two types of forces[1,2,13,14]: electrostatic forces (weak) and those generated by thermal expansion (slow). Small electromagnetic motors cannot generate forces of useful magnitude but fast and strong piezoelectric elements are not compatible with microtechnology and need a high voltage for actuation. Some progress was achieved with the use of electroactive polymers[15-17] but they are not suitable for many applications in microsystems. Electrochemical actuation was also discussed in many papers[18-24] but it is notoriously slow. One can produce a large amount of gas in a short time but it is impossible to get rid of this gas also fast. In this paper we present an actuator that uses alternating polarity (AP) short-time electrolysis[6,7] to produce $H_2$ and $O_2$ gases but termination of the gases happens very fast due to spontaneous combustion.

**Results**

**Fabrication and characterization.** The actuators were fabricated on Si wafers covered with a layer of silicon rich nitride (SiRN) (thickness 530 nm) containing deposited Pt electrodes of different designs. This layer played the role of the actuator membrane. The membrane was made free by etching the wafer from the back side. The chamber and filling channels were isotropically etched in borofloat wafers. The glass and Si wafers were anodically bonded as shown in Fig. 1. A polysilicon thermal sensor designed for four-probe measurements was fabricated underneath the electrodes (see details in Supplementary). The bonded wafers were diced into separate chips with the size 2×1 cm$^2$ . The actuator chamber with the nominal dimensions 100×100×5 µm$^3$ was filled via the channel (cross section 10×5 µm$^2$) with a solution of $Na_2SO_4$ in deionized water. In most cases the salt concentration was around 1 M. The inlet and outlet openings were sealed with a foil, the chip was glued to a printed circuit board (PCB) with an opening in the middle to get access to the membrane from the back side. For one chip without the covering glass we calibrated the membrane deflection $d$ in its center versus applied external pressure $\Delta P$. It was found that $\Delta P = Ad + Bd^3$ where $A = 2.03$ bar/µm and $B = 0.27$ bar/µm$^3$. This relation can be used to estimate the pressure in the chamber if $d$ is known.

Short square voltage pulses were applied to the electrodes. The response of the system was observed with a homemade stroboscope[25] and with a vibrometer (Polytec MSA-400). In the latter case it was possible to observe the process from the top (via glass) or from the bottom focusing the laser beam ($\lambda$ = 633 nm) with a diameter of 1.5 µm on the membrane. First, we measured the membrane deflection due to applied voltage pulses from the back side. For this configuration the change of the refractive index of the liquid due to different reasons (dissolved gas, heating, pressure change) does not influence the signal of the interferometer. Bubbles with the radius $r > \lambda/2\pi$ appearing in the chamber can scatter the light because the membrane is transparent. To eliminate their effect on the signal we focused the laser beam on the opaque electrode. With these precautions we can be sure that the measured signal corresponds to the velocity of the membrane about 10 µm off its center (see Supplementary Fig. S1).

4**Electrochemical cell.** To achieve a significant actuation stroke in a short time we have to produce a large amount of gas. For this a large current through the chamber is needed. An electrochemical cell supports a large current in the ohmic regime when the current $I$ and voltage $U$ are related as[26] $I \approx (U - E_0)/R$. Here $E_0$ is the water dissociation potential and $R$ is the resistance of the cell. Both of the parameters are defined by the cell and it was already noted[7] that for microsystems these parameters are considerably larger than for macrocells. For example, the $I - U$ curve for one of our sample (see Supplementary Fig. S2) corresponds to $E_0 = 2.8$ V and $R = 3.7$ kΩ. Therefore, to run the actuation fast we need a larger voltage than usually used for electrolysis. In our experiments we used AP pulses with the amplitude $U = 5 - 10$ V.

The average current density for our chips is around 200 A cm$^{-2}$. It results in a very high relative supersaturation $S > 1000$ nearby the electrode surface. When $S$ is so large the bubbles nucleate homogeneously and exist in the form of nanobubbles during 100 µs or so[7]. For single polarity electrolysis the bubbles finally grow to a well visible microscopic size, but for AP electrolysis only small number of microbubbles is observed if the switching frequency is larger than 20 kHz (see[6] and Supplementary Fig. S3). Some additional information and estimates are given in Supplementary Information. Let us note that for macroscopic electrolytic cells observed on a long-time scale[27] the maximal current density is ~1 A cm$^{-2}$ and maximal supersaturation is ~100.

**Membrane deflection.** For AP voltage pulses with the amplitude $U = \pm 10$ V repeated with the frequency $f = 50$ kHz (driving or switching frequency) during $\tau = 600$ µs the signal and current are shown in Fig. 2. The signal consists of separate narrow lines of increasing amplitude correlated with the driving pulses. It is well visible in Fig. 2d where only five periods for both the signal and current are shown in the same plot. The signal in Fig. 2a integrated over time $t$ is the deflection of the membrane $d(t)$. It is shown in Fig. 2c. The deflection increases with time finally reaching a steady state while oscillations with the driving frequency $f$ are superimposed on the smooth curve (see enlarged view in Fig. 2e). In its maximum the deflection becomes as high as $d \approx 1.4$ µm that corresponds to the pressure increase in the chamber $\Delta P \approx 3.6$ bar.

When the electrochemical process is switched off the pressure drops very fast (see Fig. 2c). This is our key observation that opens the way for a closed actuation cycle. A huge amount of gas disappears in 100 µs or so. The maximal deflection of the membrane corresponds to the apparent volume of gas $\Delta V = 0.7 \times 10^4$ µm$^3$ at the pressure $P \approx 4.6$ bar. As the pressure is reduced back to normal this gas has to fill 60% of the chamber but no gas is visible. The actual amount of gas is much larger. This is because most of the gas is densely packed in nanobubbles (see[6,7] and estimates in Supplementary Information). Only small amount of gas can be dissolved in the chamber to saturate the liquid. Diffusion out of the chamber via the filling channels is negligible on this timescale. Moreover, the gas produced by the single polarity electrolysis, when hydrogen and oxygen are separated in space (appear near different electrodes), is well visible and exists on a longer timescale (see Supplementary Fig. S3). The only reasonable explanation of the observed fast pressure decrease is consumption of the gases in the overall reaction of water formation. We cannot separate elementary steps in this reaction but the process is not catalytic. In 600 µs the gases diffuse too far away from the electrodes (a few µm) to be consumed in surface reactions in 100 µs.

**Discussion**

Our interpretation of the physical events presented in Fig. 2 is the following. Stoichiometric nanobubbles that are formed nearby the electrode surface in phase with the electrical pulses[6] are responsible for the sharp peaks in Fig. 2a and 2d. Previously we observed periodic reduction of gas concentration in the electrolyte above the electrodes[6]. Here we can see that in the closed chamber the process is accompanied also by the periodic variation of the pressure. This is an independent argument that the combustion reaction happens in the nanobubbles. The monotonic pressure increase in Fig. 2c is due to the gas existing as separate hydrogen and oxygen nanobubbles and dissolved molecules. The nanobubbles are formed homogeneously and very fast because the local supersaturations is high. For each period more gas is produced than consumed while a steady state is not reached. In the steady state a part of the produced gas disappears in correlation with the driving pulses (high frequency oscillation). The rest of gas gets into the reaction independently on the electrical pulses (randomly) by formation of stoichiometric nanobubbles in the solution. This random process is responsible for the fast decrease of the pressure after the switch off. The details of this interpretation can change, but the solid fact – the fast pressure decrease – has to be related to the reaction between $H_2$ and $O_2$.

*Thermal effect.* The deflection of the membrane demonstrates dependence on the driving frequency *f* as shown in Fig. 3a. The higher the frequency the smaller amount of gas escapes the reaction resulting in a smaller increase of pressure. For this reason the heat produced by the combustion reaction must increase with *f*. On the contrary, the Joule heating does not depend on *f*. This is a clear experimental signature to distinct between the two sources of heating. To see the thermal effect we observe variations of the electrolyte conductivity similar to that for any resistive sensor. We fit each current pulse (half of a period) with the function[7] $I(t) = I_F + I_1 e^{-kt}$, where $I_F$ is the Faraday current and the second term (parameters $I_1$ and $k$) describes surface charge-discharge processes[28]. Figure 3b shows the Faraday current as a function of time for different frequencies. The current increase with time demonstrates the effect of heating. Faster increase for higher frequencies shows that at least a part of the total heat is produced by the reaction. Observation of the heating became possible due to small thermal mass of the membrane.

To estimate an effective temperature in the chamber we measured (using METTLER TOLEDO SevenMulti) the conductivity of the bulk solution, $\sigma = \sigma_0(1+a\Delta T)$, as a function of the temperature increase $\Delta T$ to find $\sigma_0 = 15$ $\Omega^{-1}m^{-1}$ at $T = 20$ °C and $a = 0.024$ $K^{-1}$. Independently $a = 0.024\pm0.001$ $K^{-1}$ was determined from our samples using external heating. The temperature dependence of the Faraday current was used to extract information on the effective temperature (see inset in Fig. 3b). The temperature determined in this way is closer to the maximal temperature than to the average one because the most significant contribution to the current comes from the hottest regions around the electrodes. Our built-in thermal sensors were not fast enough to measure the temperature change on the time scale 100 µs presumably due to parasitic electrical effects.

Even for very high supersaturation the homogeneous nucleation is still an activation process. Heating increases the nucleation rate including the bubbles containing only $H_2$ or $O_2$ gases. Because more unreacted gas appears in the chamber the pressure has to increase faster. This effect is responsible for faster than linear increase of $d(t)$ in Figure 3a. A similar behavior has to be observed when the external temperature increases. To see this effect one chip was glued to a flat resistive heater and calibrated with a thermocouple. This configuration was observed from the top. The vibrometer results were equivalent to that observed from the bottom except of occasional scattering on microbubbles appearing more often at low frequencies or high currents. The results shown in Fig. 4a demonstrate significant dependence of the deflection on temperature. Scattering on microbubbles is visible as some irregularity for the curve at $T$= 30 ˚C.

*Actuation frequency.* The dynamics of cooling was observed by applying two series of driving pulses separated by a delay time. The membrane deflection is shown in Figure 4b. The first series of pulses 100 µs long heats up the system. The second series is started when the temperature nearby the electrodes is augmented resulting in a larger deflection than the first one. When the delay between the series of pulses increases the effect fades away. The same experiment demonstrates how fast the membrane can be actuated using series of pulses separated in time (see also Supplementary Fig. S4). From our data it follows that cyclic operation with an actuation frequency of $F$ = 5 kHz is feasible and at this frequency the actuator is able to deliver the overpressure on the level of 1 bar. Small increase in the external temperature can increase both the actuation frequency and the developed force. The maximal stoke can be reached for the minimal driving frequency around $f$ = 20 kHz, however, for small $f$ degradation of electrodes is observed[6] and the optimal driving frequency is a matter of durability and material choice. Of course, the actuation with the driving frequency $F = f \sim 100$ kHz is also possible but with a smaller amplitude (pressure).

In conclusion, the results reported in this paper not only demonstrate a fast and strong actuator that can be applied in microfluidics, micro/nano positioning, or in compact sound/ultrasound emitters. More importantly, they demonstrate feasibility of combustion reactions in microscopic volumes. This is a fundamental statement that opens up new possibilities to power micro and mini systems.

**Acknowledgements**

Funding was provided by the Dutch Technology Foundation (STW). We thank D. Lohse for numerous discussions, J. W. Berenschot for discussion of fabrication issues, and H. van Wolferen for technical assistance.

**Author Contributions**

V. B. S. performed experiments and analysis; R. G. P. S. contributed to measurements and development of the setup; K. M. contributed to the device fabrication; M. C. E. contributed to the analysis; both V. B. S. and M. C. E. contributed to the writing of the paper. All authors were involved in discussions.

**Additional information**

**Supplementary information** accompanies this paper at

http://www.nature.com/scientificreports

**Competing financial interests:** The authors declare no competing financial interests.

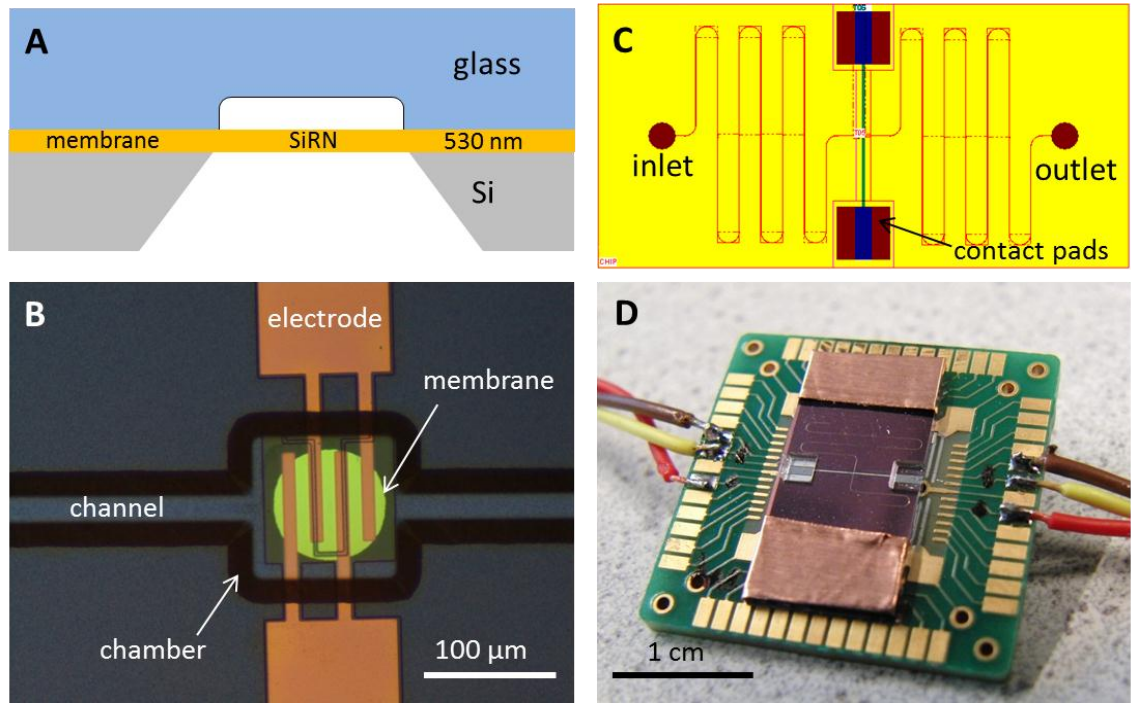

**Figure1 | Design of the chip. a**, A silicon wafer with a SiRN layer on top is bonded with a structured glass wafer. **b**, Optical image of the chamber in the device before filling. The membrane (highlighted with green light from the bottom) is not completely etched, but normally it coincides with the size of the chamber. Under the central electrodes a thermal sensor (polysilicon) is visible. **c**, General design of the chip: in/outlets, long channels, and six contact pads (2 for the electrodes and 4 for the sensor). **d**, Completely functioning device glued to a PCB, sealed and wire bonded.

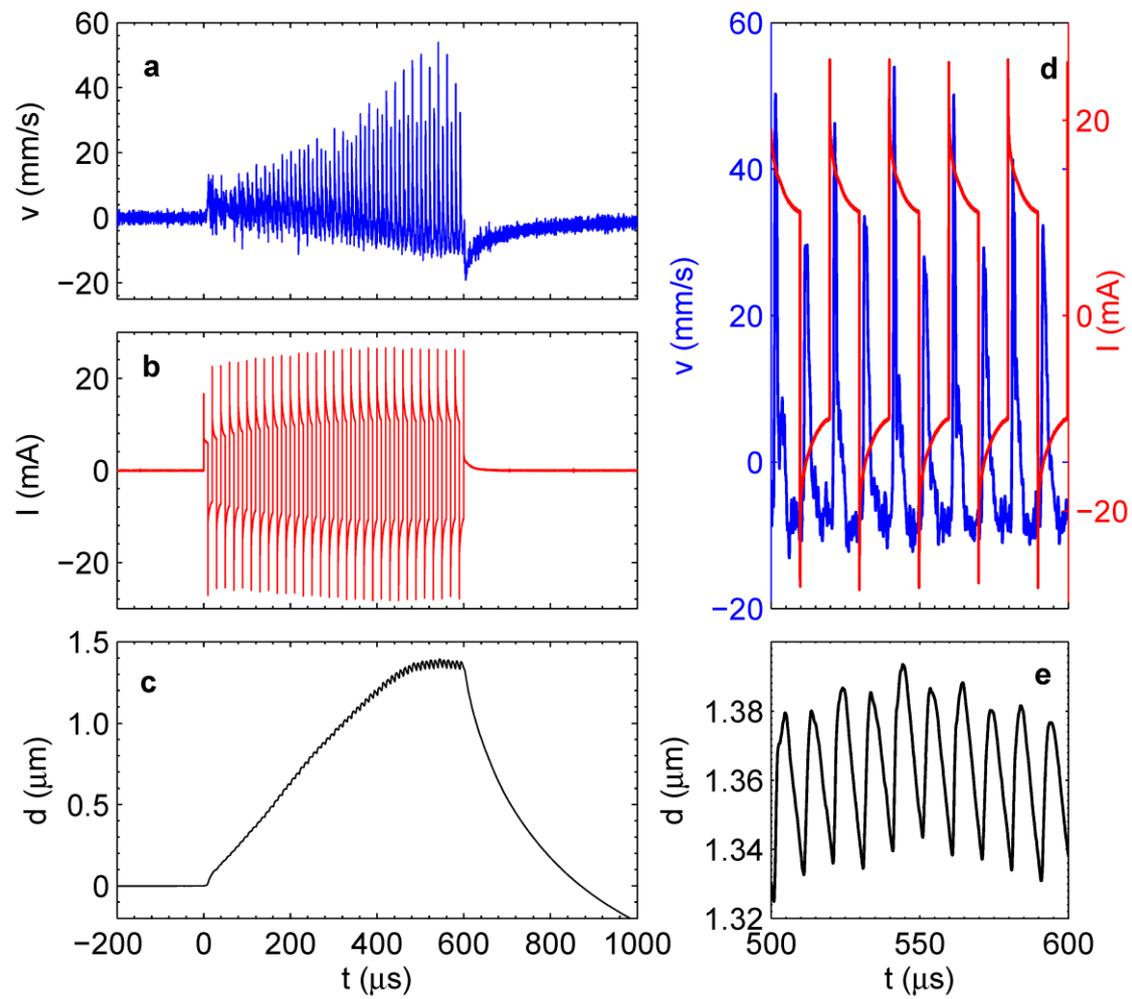

**Figure 2 | Deflection of the actuator membrane measured from the back side.**
The membrane is driven by AP voltage pulses $U = \pm 10$ V, $f = 50$ kHz, and $\tau = 600$ μs.
**a**, Velocity of the membrane (raw signal of the vibrometer). **b**, Current flowing through the electrodes. **c**, Deflection of the membrane (raw signal integrated over time). **d**, Five last periods of the signal in **a** and the current in **b** are shown in the same plot. **e**, Five last periods in the membrane deflection.

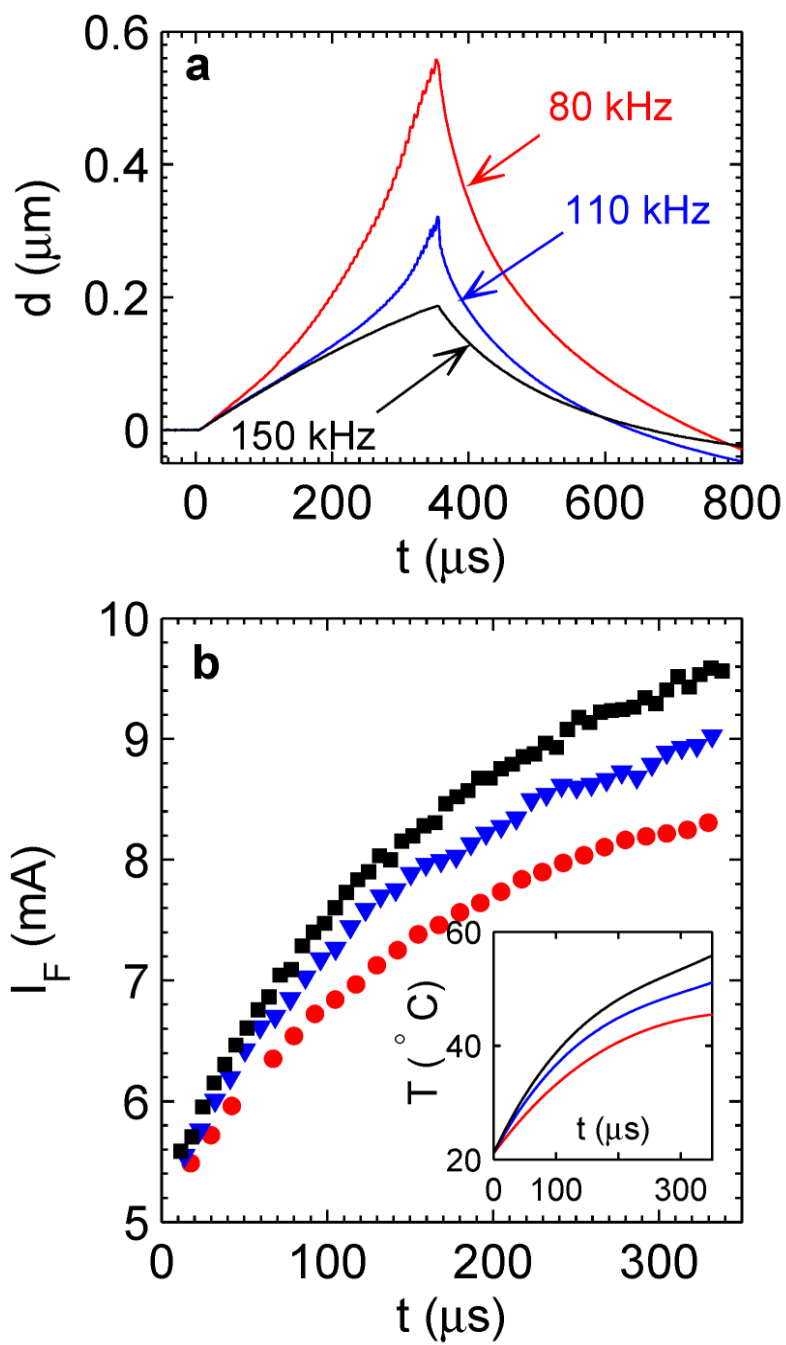

**Figure 3** | **Heating due to combustion of gases. a**, Deflection of the membrane for different driving frequencies ($U$ = ±9 V, τ = 350 µs). **b**, The Faraday current as a function of time for the runs presented in **a** (one point per period). The inset shows the effective temperature (smoothed) in the chamber. Different colors correspond to the frequencies shown in **a**.

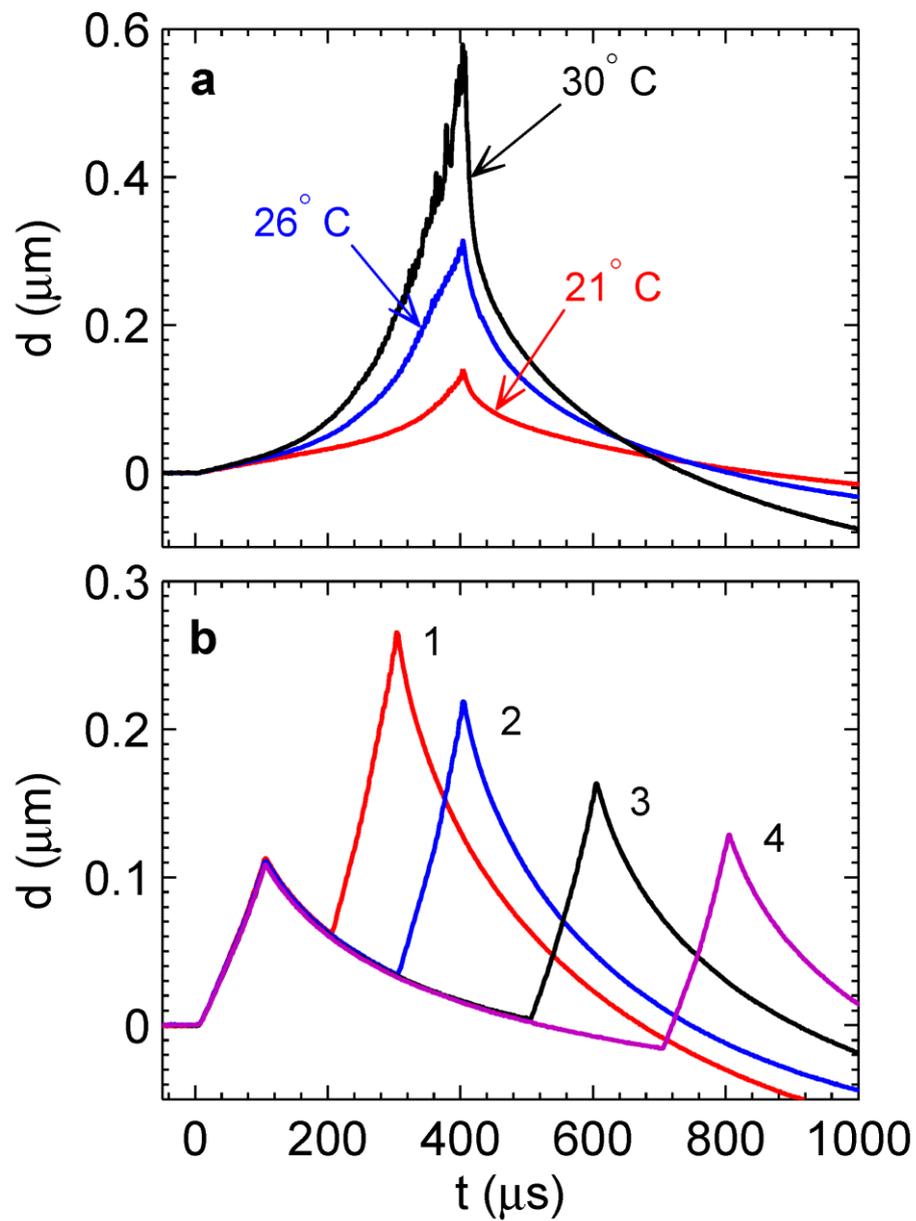

**Figure 4 | Influence of heating on the actuator performance.** **a**, Membrane deflection for different external temperatures ($U = \pm 10$ V, $f = 100$ kHz, $\tau = 400$ μs, low current sample). **b**, Membrane deflection for two successive series of pulses 100 μ*s* long each separated by different delay times. For the curves 1, 2, 3, and 4 the delay is 100, 200, 400, 600 μs, respectively.

# Supplementary Information

# New type of microengine using internal combustion of hydrogen and oxygen

Vitaly B. Svetovoy[*,1,2], Remco G. P. Sanders[1], Kechun Ma[1] & Miko C. Elwenspoek[1,3]

[1]MESA+ Institute for Nanotechnology, University of Twente, PO 217, 7500 AE Enschede, The Netherlands (v.svetovoy@utwente.nl)

[2] Institute of Physics and Technology, Yaroslavl Branch, Russian Academy of Sciences, 150007,Yaroslavl, Russia

[3] FRIAS, University of Freiburg, 79104 Freiburg, Germany

**Supplementry Figures**

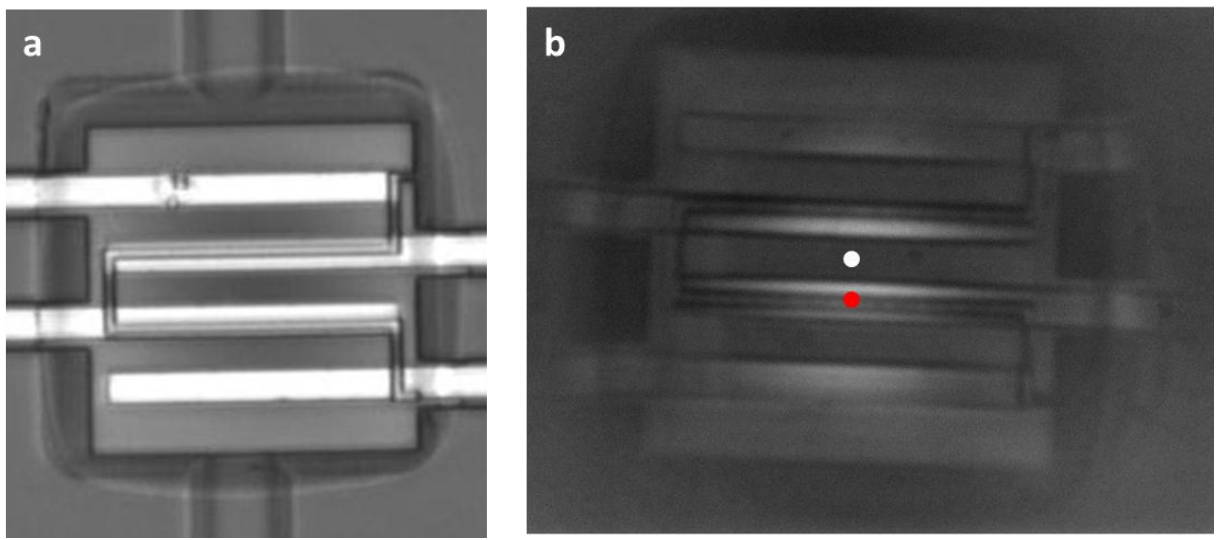

**Figure S1. Optical images of the filled chamber made with the vibrometer. a**, Top view through the glass wafer. **b**, Bottom view. The membrane is buried 380 μm deep in the silicon wafer. A long-focus objective of the vibrometer is used for the image. The red circle shows position of the laser beam on the opaque electrode. The white circle indicates the center of the membrane.

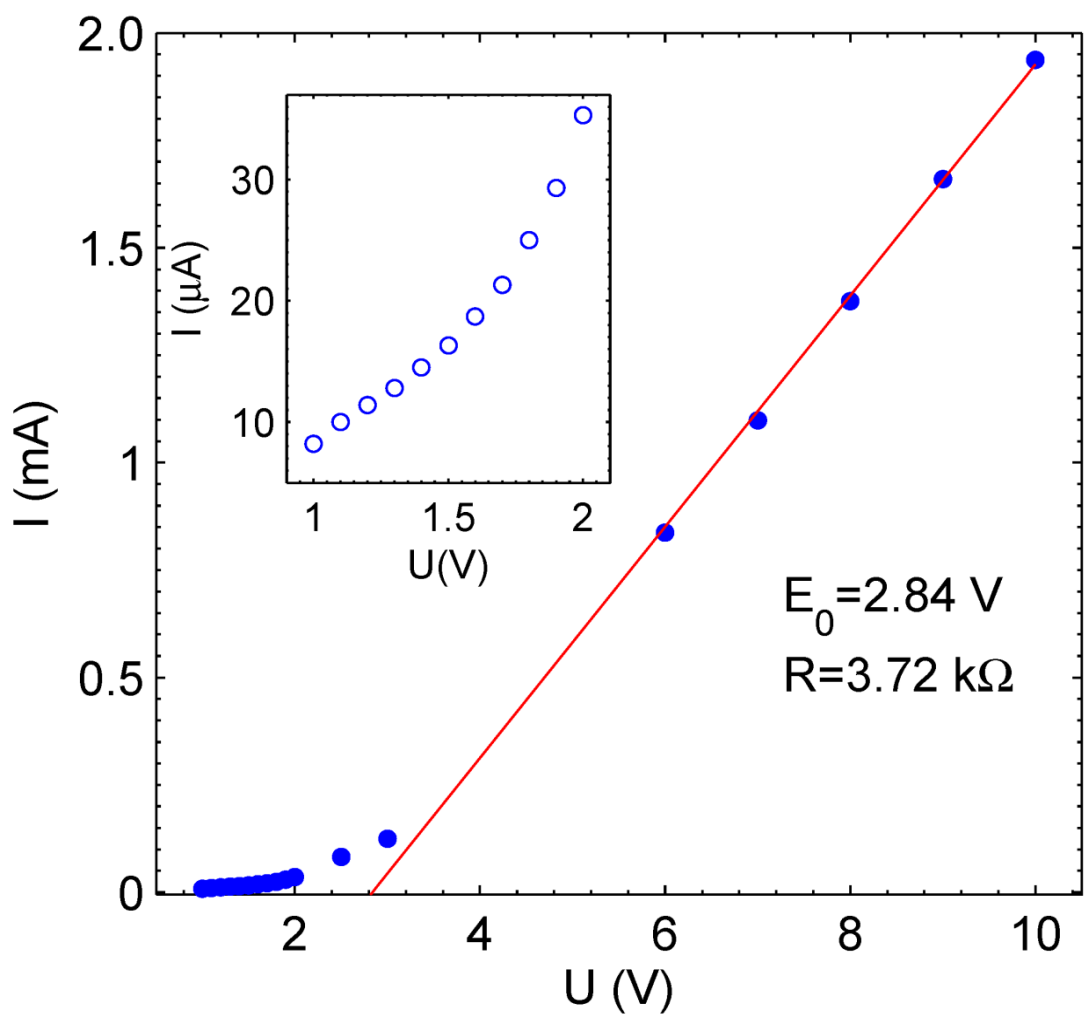

**Figure S2. Current-voltage characteristics** of the microscopic electrochemical cell for one of the samples (low current sample). At $U > 3$ V it was not possible to neglect the heating by the current. The corresponding points were collected using the procedure similar to that used for Fig. 3b. Then the current was extrapolated linearly to $t = 0$ to exclude the effect of heating. The red line is the best fit with the function $I = (U-E_0)/R$, where the parameters $E_0$ and $R$ were found from this fit. The inset shows enlarged view for the points below 2 V.

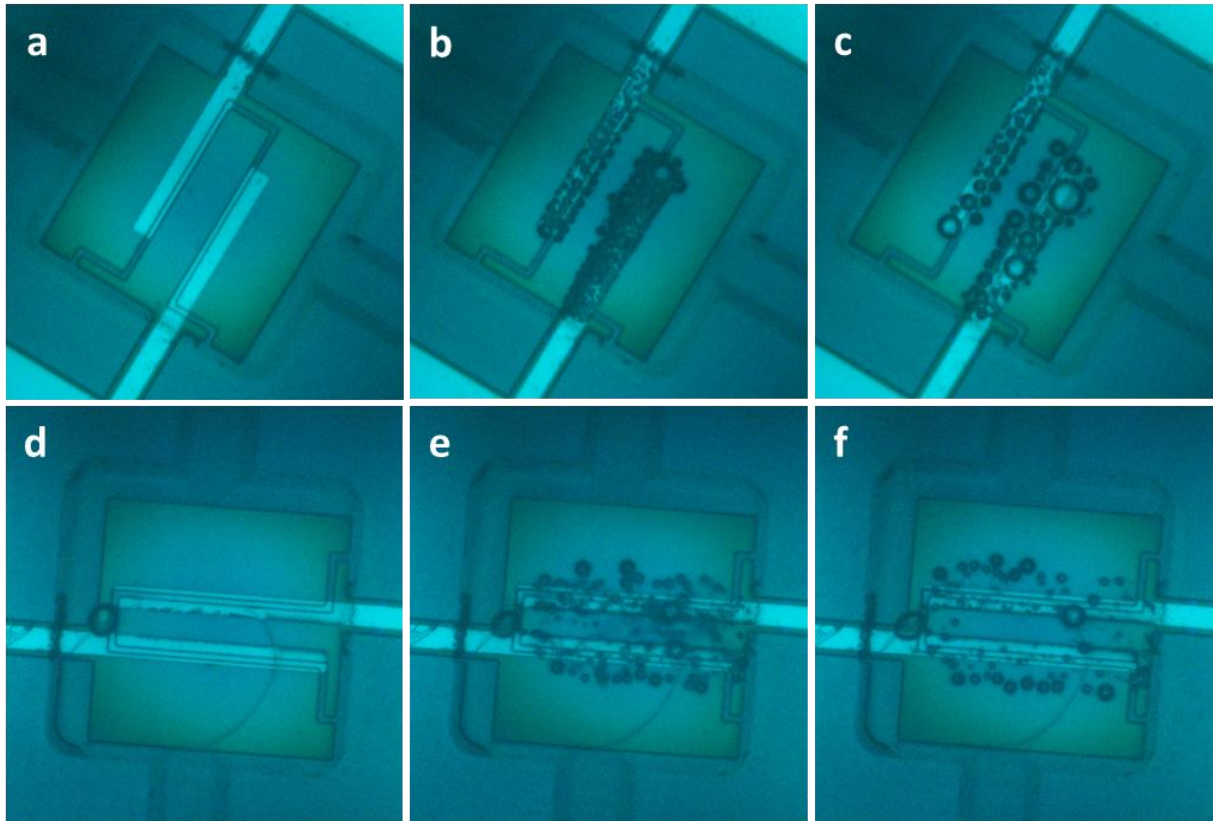

**Figure S3. Stroboscopic images of the chamber** (flash is 10 µs long). The top row is for single polarity pulse with $U$ = 8 V, $\tau$ = 0.2 ms. The panels **a**, **b**, and **c** correspond to the observation time $t$ = 0, 0.2, 2 ms, respectively. The gas forms well visible microbubbles that coalesce with time and diffuse slowly. The bottom row is for AP pulses with $U$ = ±8 V, $f$ = 100 kHz, $\tau$ = 0.8 ms. The panels **d**, **e**, and **f** correspond to $t$ =0, 0.8, 1.6 ms, respectively. In the bottom row the voltage was applied longer and the current was nearly 3 times higher (different sample) than in the top row. Nevertheless, the number of microbubbles is smaller for AP pulses and they have different pattern. The number of visible microbubble decreases further with the increase of the driving frequency. Nanobubbles can be seen in the image **e** as a weak haze and they are on the edge of visibility.

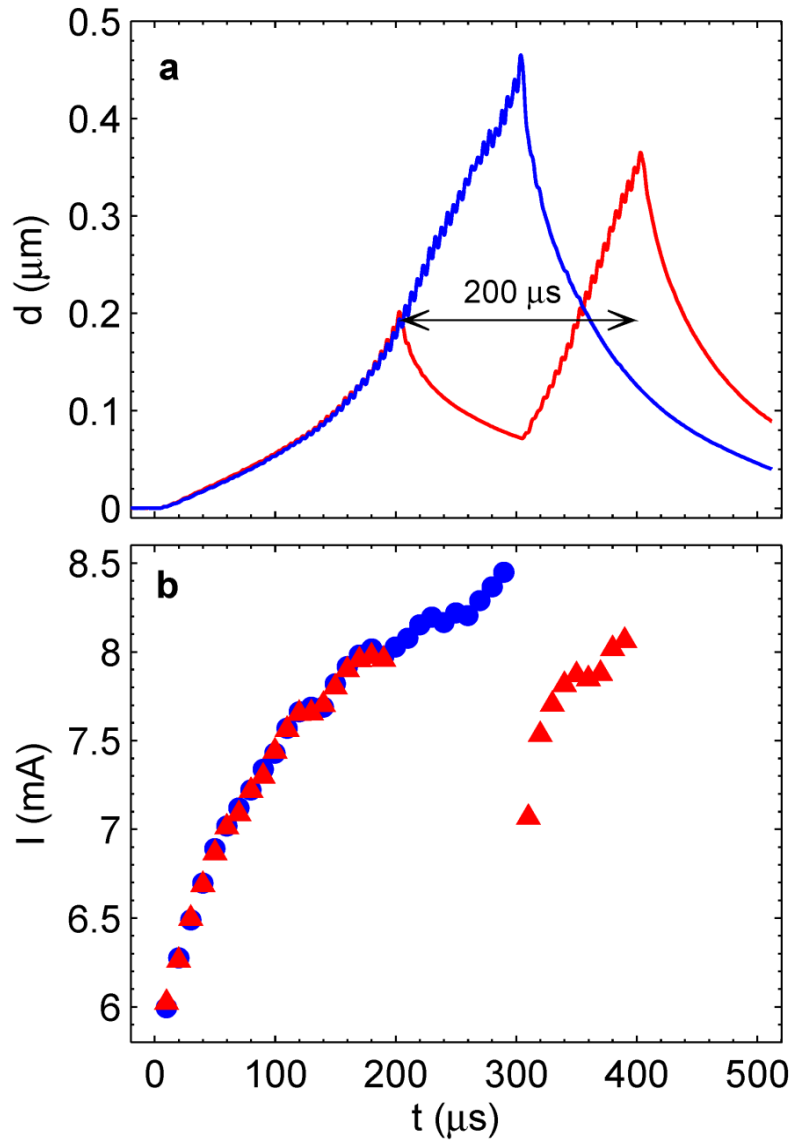

**Figure S4. Continuous series of pulses vs two series separated by a delay.** The runs were for $U = \pm 9$ V, $f =100$ kHz and the chamber was observed from the top. **a**, Deflection of the membrane. The blue curve corresponds to 300 μs long continuous series of pulses. The red curve is the response on the two series of pulses 200 μs and 100 μs long separated by 100 μs delay. The red curve demonstrates possibility of actuation with a frequency of $F = 5$ kHz. **b**, Faraday current for the runs given in **a** (one point per period). The blue circles are for the continuous series. The red triangles are for the two series of pulses separated by a gap of 100 μs long.

**Details of fabrication**

In our design the electrodes and the thermal sensor underneath the electrodes are fabricated directly on the membrane. Here we provide some details of fabrication. Silicon wafer was covered with 530 nm of SiRN. Half of micrometer of polysilicon was deposited on top of the nitride, which was doped with boron to a resistivity of around $2\times10^{-5}$ Ω·m. The polysilicon layer was patterned to form a resistor as shown in Fig. S5a using the standard lithography process and reactive ion etching. For insulation of polysilicon from the electrodes and for good bonding with the glass wafer the top layer (170 nm) of polysilicon was oxidized. The electrodes were deposited by sputtering. A layer of Ti 10 nm thick was deposited for better adhesion followed by 100 nm of platinum. For patterning of metals the lift-off process has been used. The resulting structure on the membrane is shown schematically in Fig. S5b.

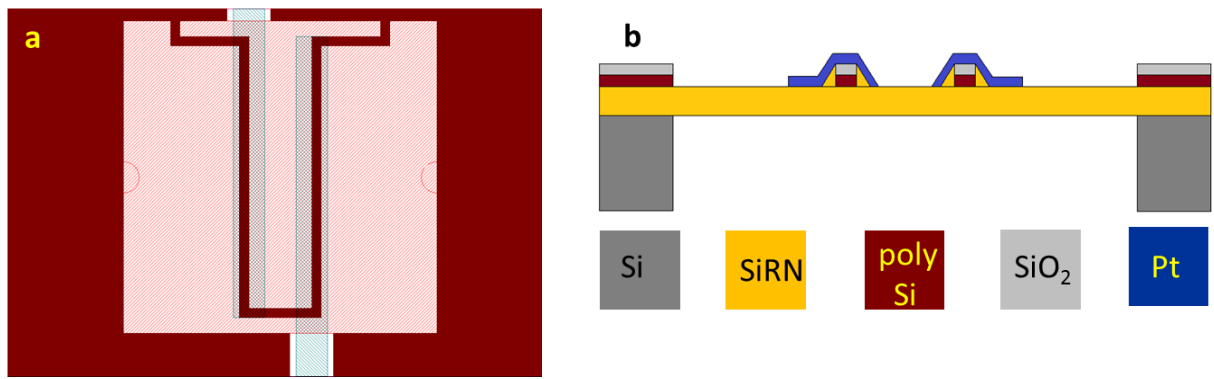

**Figure S5. Design of the thermal sensor. a**, Polysilicon layer (top view). **b**, Membrane, electrodes, and polysilicon sensor (cross section).

To cover the polysilicon resistor smoothly without cracks before the sputtering we deposited SiRN and wet etched it isotropically. Nitride left in the inner corners and guaranteed smooth covering with metals as shown in Fig. S6.

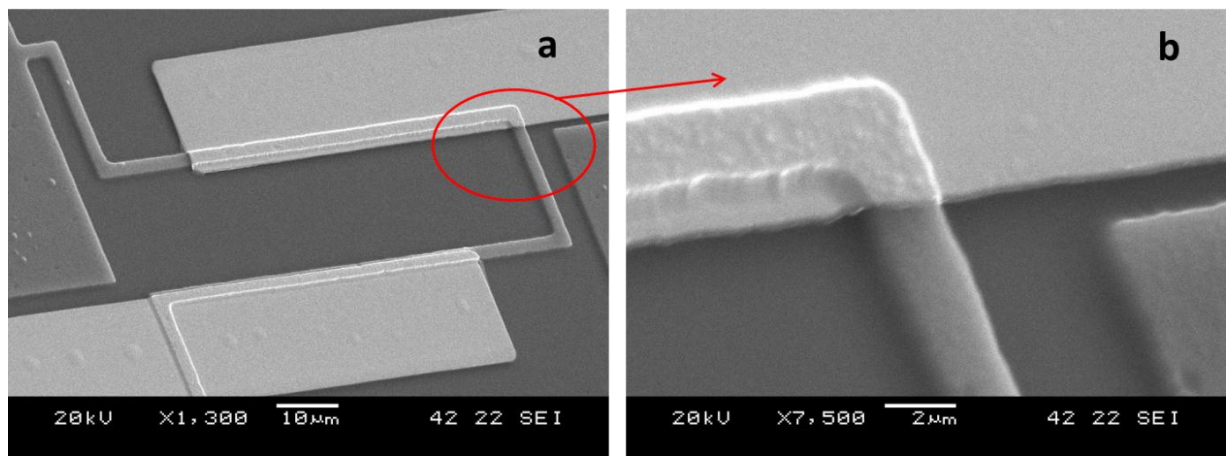

**Figure S6. Covering of polisilicon resistor with Pt electrodes. a**, SEM image of the wire resistor and electrodes. **b**, Magnified image of the circled area demonstrates smooth coverage of the polisilicon wire.

**Supplementary Text**

Let us estimate roughly how much gas exists in the chamber using as an example the data in Fig. 2. The total amount of gas molecules $N$ produced by the electrochemical process is given by the expression

$$N = \frac{3}{4e}\int_0^\tau dt\, I_F(t), \qquad (S1)$$

where $e$ is the absolute value of the electron charge and the Faraday current $I_F(t)$ is understood as a continuous function of time in the same sense as presented in Fig. 3b. Extracting $I_F(t)$ from the data in Fig. 2b and taking the integral for τ = 600 μs we find $N \approx 2.5 \times 10^{13}$.

Let us assume for the moment that the gases cannot react and all the produced gas will stay in the system. At normal pressure this gas would fill $20V_0$, where $V_0 = 5 \times 10^4$ μm$^3$ is the nominal volume of the chamber. This huge amount of gas is not observed and has to disappear in some way.

Significant part of the produced gas disappears via stoichiometric nanobubbles formed nearby the electrode surface. This process is well documented in[6,7] but for completeness let us repeat and refine the physical arguments. The first pulse (negative for definiteness) with the length $T/2 = 1/2f$ = 10 μs produces H$_2$ molecules above the working electrode. During the time $T/2$ the molecules can diffuse away from the electrode on the distance $l_D \approx \sqrt{DT/2}$, where $D = 4.5 \times 10^{-9}$ m$^2$ s$^{-1}$ is the diffusion coefficient of hydrogen in water. Numerically one has $l_D$ = 210 nm. The number of molecules produced by the first pulse is

$$N_{H_2} = \frac{I_F T}{4e}, \qquad (S2)$$

where $I_F$ = 6 mA is the current in the first pulse. The density of gas molecules in the diffusion layer of thickness $l_D$ is

$$n_{H_2} = \frac{N_{H_2}}{Al_D} \approx 5 \times 10^{20} \, \text{cm}^{-3}, \tag{S3}$$

where $A$ = 2000 µm² is the area of the working electrode. Comparing this estimate with the saturated concentration of hydrogen $4.7 \times 10^{17}$ cm⁻³ one finds the relative supersaturation $S$ = 1000 averaged over the electrode surface. Note that the pressure at the first pulse is still close to the atmospheric pressure. The current density is distributed nonhomogeneously and locally $S$ reaches even higher values. With this level of supersaturation the bubbles nucleate homogeneously and very fast. The upper limit on the bubble size is $2r < l_D$ = 210 nm, where $r$ is the bubble radius. These small bubbles cannot be observed optically because they scatter light very weakly.

The next pulse is positive and it produces oxygen above the working electrode. For oxygen one finds similar but somewhat smaller values: $l_D$ = 140 nm and $S$ = 430. Oxygen bubbles also can be nucleated homogeneously but with high probability $O_2$ molecules will diffuse in the existing nanobubbles containing hydrogen. In this way stoichiometric nanobubbles can be formed. In these bubbles the reaction happens spontaneously[6] but the mechanism is still unclear. Disappearance of the stoichiometric nanobubbles in phase with driving pulses is supported by observation of the gas density oscillations in Ref. [6] and the membrane oscillations in this study, among other arguments.

We can estimate the number of gas molecules in the chamber $N_{ss}$ when the steady state is reached. Let us assume that all gas is collected in nanobubbles of maximal radius $r$ = 100 nm (larger bubbles would be visible due to significant light scattering). This gas can be described by the gas law

$$(P_0 + \Delta P + P_L)\Delta V = N_{ss}kT, \tag{S4}$$

where $P_0$ is the atmospheric pressure, $\Delta P$ = 3.6 bar is the overpressure in the chamber, and $P_L = 2\gamma/r$ = 14 bar is the Laplace pressure for a bubble in water. The increment of the volume is estimated for the square 100×100 µm² membrane with maximal deflection $d_0$ = 1.5 µm as $\Delta V = 0.7 \times 10^4$ µm³. Since we took the largest radius for the bubbles, Eq. (S4) gives the estimate for the smallest number of gas molecules in the steady state, $N_{ss} > 3.1 \times 10^{12}$. According to Fig. 2c the pressure in the chamber is relaxed for 100 μs or so and without reaction all this gas would fill $2.6V_0$. Because the gas is not observed we conclude that it is consumed in the reaction of water formation.